# Real-time Earthquake Early Warning with Deep Learning: Application to the 2016 Central Apennines, Italy Earthquake Sequence

*Xiong Zhang[1], Miao Zhang[2*,3], Xiao Tian[1]*

*[1]Fundamental Science on Radioactive Geology and Exploration Technology Laboratory, East China University of Technology, Nanchang, Jiangxi 330013, China*

*[2]Department of Earth and Environmental Sciences, Dalhousie University, Halifax, Nova Scotia B3H 4J1, Canada*

*[3]Department of Geophysics, Stanford University, Stanford, California 94305, USA*

**Abstract:** Earthquake early warning systems are required to report earthquake locations and magnitudes as quickly as possible before the damaging S wave arrival to mitigate seismic hazards. Deep learning techniques provide potential for extracting earthquake source information from full seismic waveforms instead of seismic phase picks. We developed a novel deep learning earthquake early warning system that utilizes fully convolutional networks to simultaneously detect earthquakes and estimate their source parameters from continuous seismic waveform streams. The system determines earthquake location and magnitude as soon as one station receives earthquake signals and evolutionarily improves the solutions by receiving continuous data. We apply the system to the 2016 Mw 6.0 earthquake in Central Apennines, Italy and its subsequent sequence. Earthquake locations and magnitudes can be reliably determined as early as four seconds after the earliest P phase, with mean error ranges of 6.8–3.7 km and 0.31–0.23, respectively.

## INTRODUCTION

Earthquake early warning (EEW) is a practical way to mitigate seismic hazards by providing source parameters prior to significant ground shaking (*1*). The warning time depends on the distance between the user and earthquake epicenter, the EEW algorithm, and the station distribution. Epicentral regions suffer from the largest damage but have the least time and constraints to early warning alarms. Deep learning techniques provide potential for extracting earthquake source information from full waveforms, potentially making it possible to utilize fewer earthquake signals to reduce the early warning response time. We developed a real-time EEW system by directly mapping full waveforms to source parameters using a fully convolutional network that was originally developed for image segmentation in computer vision.

### EEW algorithm review

EEW systems aim to rapidly and reliably provide earthquake source information before the damaging S wave arrival (*2-4*). They detect hazardous earthquakes, estimate their source parameters, and transmit warnings to the public. Conventional EEW algorithms depend on picking and analyzing P phases and are generally divided into two categories: onsite warning and regional warning (*1, 4*). For onsite warning, predominant periods and/or amplitudes of the first few seconds of P waves at single stations are utilized to evaluate source magnitude or ground shaking (*5-10*). For regional warning, arrival times and amplitudes of P waves at multiple stations are used to estimate earthquake locations and magnitudes, providing more accurate source parameter solutions and more reliable and complete warning information (*1, 4, 11, 12*). However, as pick-based methods, regional warning algorithms are composed of many steps: phase detection, phase picking, phase association, earthquake location, and magnitude estimation (*13, 14*). On the one hand, either of the above steps may potentially affect source parameter solutions. On the other hand, they require more time to receive and analyze earthquake signals at multiple stations. It turns out that early warning cannot be issued until a sufficient number of close stations are triggered, creating a large "blind zone" (*4*). In contrast, earthquakes usually cause the largest damage in epicentral regions. To eliminate such issues, methods based on the concept of "the triggered and not-yet-triggered stations" have been proposed and applied to constrain earthquake locations, because earthquakes most likely occur near triggered stations and relatively far from not-yet-triggered stations (*15, 16*). However, such methods only constrain earthquakes in specific regions—instead of providing accurate earthquake locations—and suffer from false alarms, especially if the number of triggered stations is small or some stations are malfunctioning. Thus, the development of a fully automatic real-time EEW system by directly mapping seismic waveform data to earthquake source parameters is critical.

### Utilizing full waveforms with deep learning

Compared to seismic picks, seismic waveforms contain more information and can potentially be used to estimate earthquake source parameters with the fewest possible number of stations and to promptly transmit warning information. Deep learning techniques provide opportunities for extracting and exploiting the features behind seismic waveforms (17, 18). Recent applications for seismological studies include phase picking (*19*), phase association (*20*), earthquake location (*21*), seismic discrimination (*22*), waveform denoising (*23*), and magnitude estimation (*24*). One remarkable application is to detect and locate earthquakes from continuous waveforms through earthquake classification using the convolutional neural network, resulting in an earthquake detection rate 17 times higher than that of traditional methods (*21*). However, in the study by Perol et al (2018), the exact earthquake location was not determined because the neural network is only able to classify earthquakes into groups. Additionally, an enormous number of samples would be required to train the classification network to obtain the location with precision comparable to that of traditional location methods (*25, 26*).

### Real-time EEW with a fully convolutional network

Precision issues can be addressed by analogizing the earthquake location to the image segmentation problem using a fully convolutional network (FCN) developed for image processing (27-29). Zhang et al. (2020) demonstrated that despite the small number of available training samples, the FCN method shows promise for the determination of accurate earthquake location. However, the neural network was designed for earthquake location alone and could not handle real-time monitoring and magnitude estimation, making it unsuitable for the practical application of earthquake early warning. In this study, we designed a multi-branch FCN for real-time earthquake early warning. One branch network is designed for earthquake location and the other for magnitude estimation. Earthquake source parameters are solved starting from the first station receiving effective earthquake signals. The solutions are then improved by receiving more data in an evolutionary way. As an application, we apply this EEW system to the 2016 Mw 6.0 Central Apennines, Italy mainshock and its subsequent sequence.

## RESULTS

### Data set

A series of moderate-to-large earthquakes struck Central Apennines, Italy, from August 2016 to early 2017 (we call it the 2016 earthquake sequence). The 2016 earthquake sequence resulted in 299 casualties and more than 20,000 homeless (*30*). Most damages come from the August 24, 2016 M 6.0 mainshock, which lacked distinct foreshocks and earthquake early warning. The study region is selected based on the distribution of mapped faults and historical earthquakes, focusing on a 48 × 96 km rectangular region with a depth range of 0–32 km (Fig. 1). To maintain consistency with station recordings of historical and future earthquakes, we adopted the 12 nearest permanent broadband seismic stations with a station interval of 16–32 km, which are operated by the National Institute for Geophysics and Volcanology. We collected 3,006 M > 2.5 cataloged earthquakes that occurred in the region from February 1, 2016 to September 25, 2018, with a magnitude range of 2.5–6.5 (Fig. S1). The M 6.0 mainshock and its following first 500 M > 2.5 aftershocks, which occurred from August 24 to August 31, 2016, were selected as the testing set. The remaining 2506 earthquakes are utilized for deep learning network training.

### Real-time monitoring of the M 6.0 mainshock

The August 24, 2016 M 6.0 mainshock was responsible for most of the damages during the 2016 earthquake sequence. We first focus on the M 6.0 mainshock to test our system. To simulate real-time earthquake monitoring, continuous waveforms at 12 stations are continuously input into the well-trained neural network with a 30 s time window and 0.5 s interval. When the predicted maximum location probability within the truncated time window is larger than our preset threshold, our network detects an earthquake and outputs the corresponding location and magnitude (see Methods). The system immediately responded to the M 6.0 mainshock as soon as the first station recorded a P phase (hereinafter called "first P phase") (Movie S1). The location and magnitude were continually updated and stabilized as more data was received. We show three snapshots at 4, 9, and 15 s after the first theoretical P arrival (Fig. 2). The epicentral location of the M 6.0 mainshock is preliminarily determined at 4 s with a 4.2 km error, perturbing around the target location over time. Depth changes little with time but is systemically 2–4 km deeper than the target depth. Magnitude is underestimated as M 4.8 at 4 s and increases to M 5.1 and M 5.6 at 9 s and 15 s, respectively, approaching the expected magnitude M 6.0. Overall, both the location and magnitude of the M 6.0 mainshock become acceptable 4 s after the first P phase.

### Real-time monitoring earthquakes on August 24, 2016

We systematically investigated the performance of real-time earthquake monitoring focusing on the day of the M 6.0 mainshock. To remove common events appearing in nearby truncated time windows, we only keep the event within 30 s after a triggered detection, which most reliably possesses the maximum location probability. We further select earthquakes by only keeping events with a maximum location probability larger than 0.97. Based on the above criteria, we detect 1836 earthquakes in one day and they show a consistent distribution pattern with the 1475 earthquakes in the catalog (Fig. 3). Among them, there are 1813 and 1470 events of M ≥ 1.0 in our results and the catalog, respectively. We focus on 34 M ≥ 3.5 cataloged earthquakes and compare our results with theirs for reliability and accuracy analysis. We recovered 33 of the 34 cataloged events with mean errors of epicentral location and magnitude of 4.0 km and 0.26, respectively (Fig. S2). Among the 33 common events, there are three events with origin time errors larger than two seconds: one is due to the nearest station malfunction and the other two are due to events occurred overlapping with others (Figs. S3 and S4). The missed earthquake occurred closely after the Mw 6.0 mainshock, overlaying with its coda waves (Fig. S5). The magnitudes of the M 6.0 mainshock and the largest M 5.4 aftershock are underestimated because of waveform clipping (Fig. S6). As a demonstration, we show one-hour real-time earthquake monitoring (Movie S1).

### Relatively large earthquake monitoring

Large earthquakes potentially cause more damage than small earthquakes. We focus on relatively large earthquakes to analyze their location and magnitude results. We adopted 43 M ≥ 3.5 earthquakes from the 500 testing samples. The mean epicentral location error of the 43 large earthquakes is 6.2 km at 4 s after the first P phase, reduces to 5.4 km at 9 s, and stabilizes at 4.1 km at 15 s (Fig. 4). Compared with the cataloged magnitude, our magnitude is systemically underestimated at 4 s—because of the very limited signals received at the beginning—and approaches the cataloged magnitude as more data is received. Although the number of large earthquakes in the training set is much less than that of small ones, the FCN model is still applicable for predicting the locations and magnitudes of large earthquakes.

### Source-parameter robustness analysis

To comprehensively test the performance and robustness of our system, we symmetrically analyze 500 testing samples. To simulate real-time earthquake monitoring, we randomly cut waveforms of the 500 testing samples to form various time windows with different waveform lengths at one or multiple stations. The results show that early warning could be activated as early as 3–4 s after the first P phase if only one or two stations receive earthquake signals. As more data is received, the mean errors of epicentral location, depth, and magnitude decrease from 8.8 km, 2.6 km, and 0.4 to 3.7 km, 1.5 km, and 0.24, respectively (Fig. 5). Earthquake locations are significantly improved during the first seconds but are perturbed after ~10 s (Figs. 5a, 5c, and 5e), potentially by complex wavefields such as coda waves. With more stations receiving effective signals, the mean errors of epicentral location, depth, and magnitude decrease from 11.5 km, 2.6 km and 0.52 to 5.2 km, 1.8 km, and 0.23, respectively (Figs. 5b, 5d, and 5f). The mean errors of epicentral location and magnitude are 10.9 km and 0.45 when only two stations record effective signals with all other noises and 6.8 km and 0.31 at 4 s after the first P arrival, potentially meeting earthquake early warning requirements.

**Table 1. Detailed settings for the neural network.** The repeated layers with the same setting are denoted by multiple signs, for example, Conv2D*2 means two convolution layers.

| Layer | Channels | Kernel size | Activation | Pooling size | Upsampling size |
|---|---|---|---|---|---|
| **Location branch** | | | | | |
| Conv2D*2 | 64 | (3, 3) | ReLU | - | - |
| Maxpooling2D | - | - | - | (1, 4) | - |
| Conv2D*2 | 128 | (3, 3) | ReLU | - | - |
| Maxpooling2D | - | - | - | (1, 4) | - |
| Conv2D*2 | 256 | (3, 3) | ReLU | - | - |
| Maxpooling2D | - | - | - | (1, 4) | - |
| Conv2D*2 | 512 | (3, 3) | ReLU | - | - |
| Maxpooling2D | - | - | - | (2, 2) | - |
| Conv2D*9 | 1024 | (3, 3) | ReLU | - | - |
| UpSampling2D | - | - | - | - | (4, 2) |
| Conv2D*3 | 512 | (3, 3) | ReLU | - | - |
| UpSampling2D | - | - | - | - | (2, 2) |
| Conv2D*3 | 256 | (3, 3) | ReLU | - | - |
| UpSampling2D | - | - | - | - | (2, 3) |
| Conv2D | 128 | (3, 3) | ReLU | - | - |
| Conv2D*3 | 64 | (3, 3) | ReLU | - | - |
| Conv2D | 64 | (1, 1) | Sigmoid | - | - |
| **Magnitude branch** | | | | | |
| Conv2D*2 | 64 | (3, 3) | ReLU | - | - |
| Maxpooling2D | - | - | - | (2, 4) | - |
| Conv2D*2 | 128 | (3, 3) | ReLU | - | - |
| Maxpooling2D | - | - | - | (2, 4) | - |
| Conv2D*2 | 256 | (3, 3) | ReLU | - | - |
| Maxpooling2D | - | - | - | (3, 2) | - |
| Conv2D*2 | 512 | (3, 3) | ReLU | - | - |
| Maxpooling2D | - | - | - | (1, 2) | - |
| Conv2D*5 | 1024 | (3, 3) | ReLU | - | - |
| UpSampling2D | - | - | - | - | (1, 4) |
| Conv2D*2 | 512 | (3, 3) | ReLU | - | - |
| UpSampling2D | - | - | - | - | (1, 2) |
| Conv2D*2 | 256 | (3, 3) | ReLU | - | - |
| UpSampling2D | - | - | - | - | (1, 2) |
| Conv2D*2 | 128 | (3, 3) | ReLU | - | - |
| Conv2D | 64 | (3, 3) | ReLU | - | - |
| Conv2D | 1 | (1, 1) | Sigmoid | - | - |

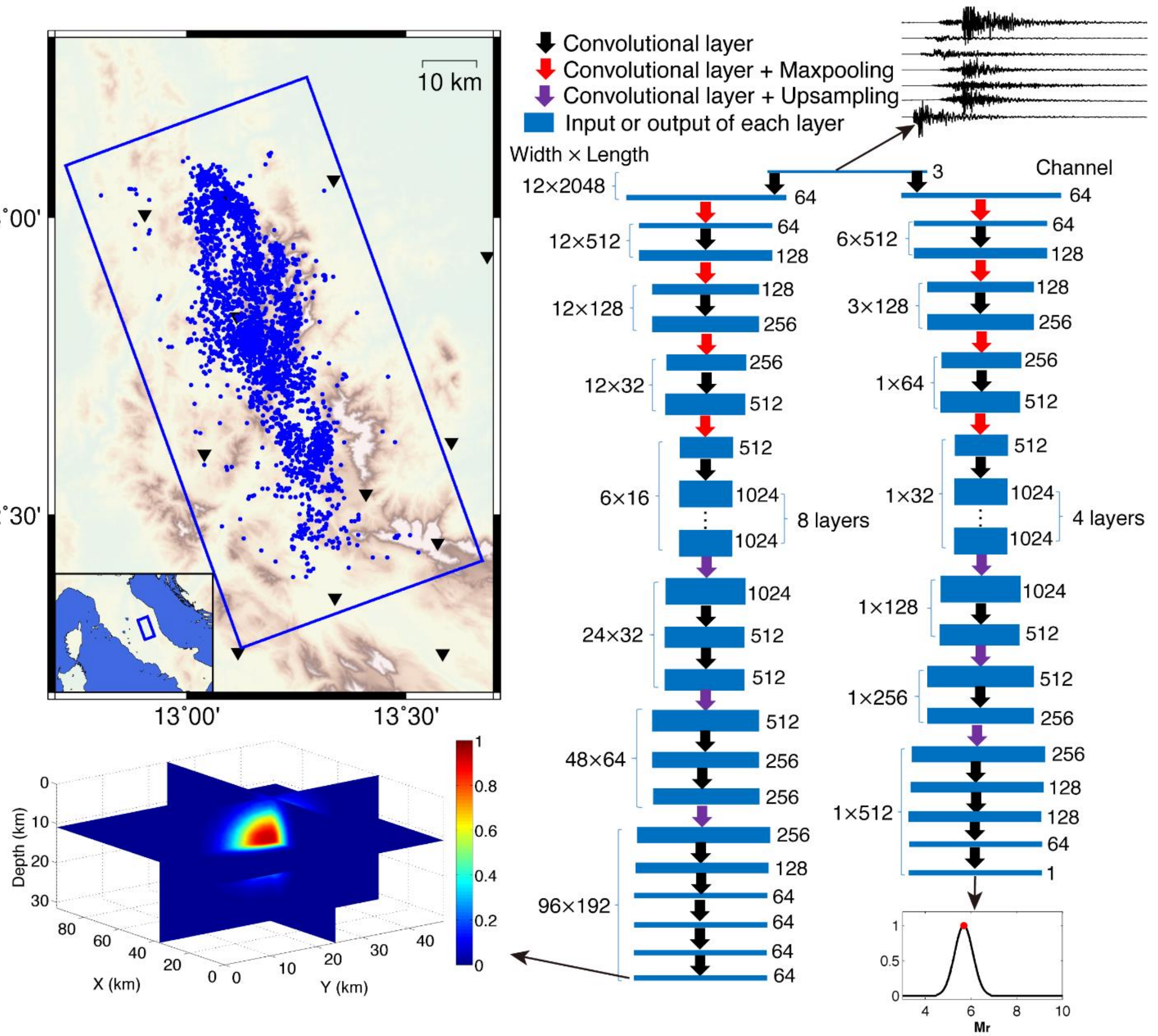

**Fig. 1**. **Map showing the study region, training dataset, and neural network architecture.** Detailed neural network settings are listed in Table 1. Earthquake location and magnitude are projected to 3D and 1D Gaussian distribution probabilities, respectively. Blue dots represent earthquakes used for neural network training; black triangles denote stations used in this study, and the rectangular box marks the study region.

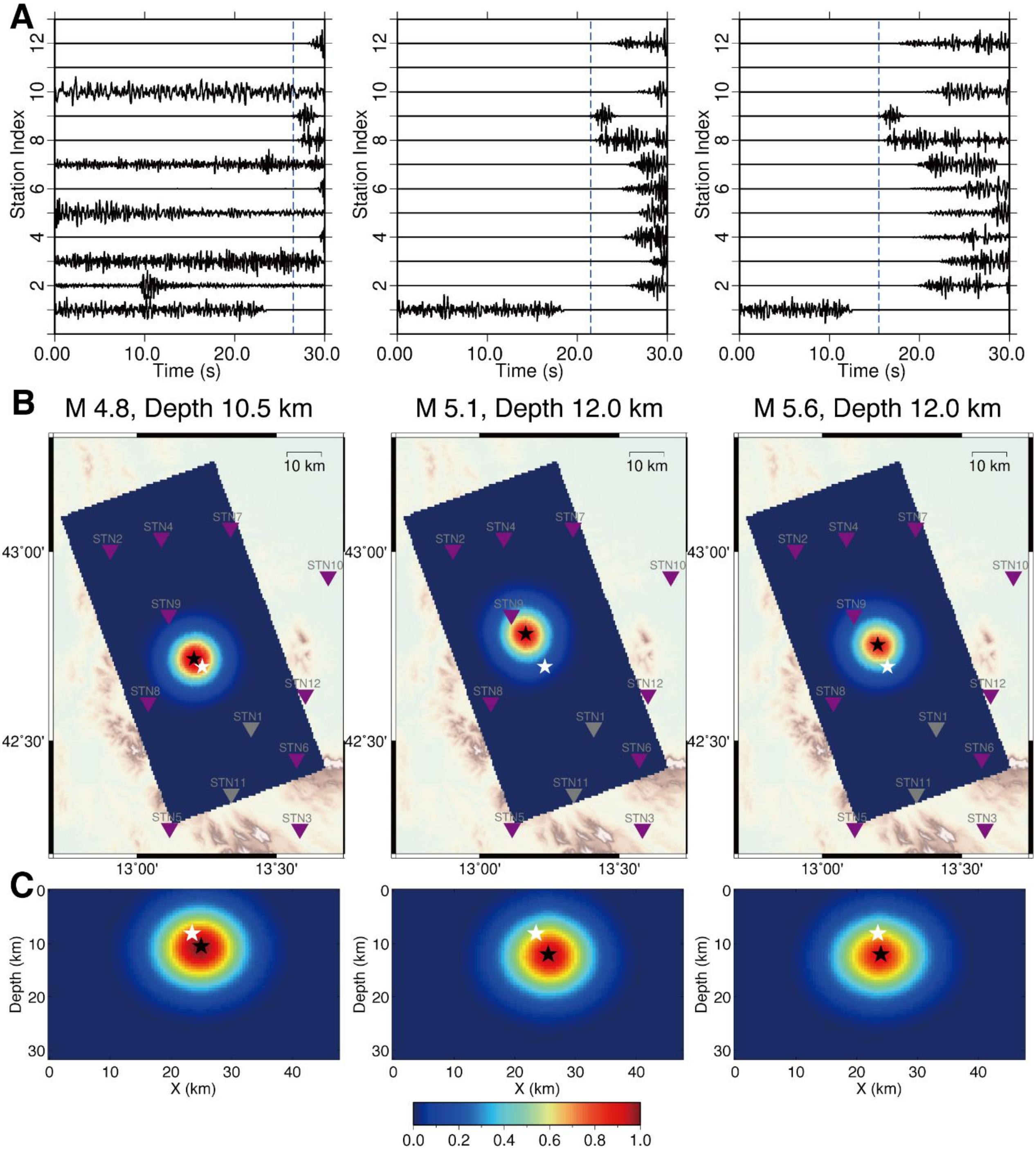

**Fig. 2**. **Snapshots of real-time monitoring of the M 6.0 mainshock.** (A) Self-normalized vertical waveforms, (B) map view, and (C) depth view of location probabilities at 4 s, 9 s, and 15 s after the first P arrival (from left to right). The dashed lines denote the first P arrival. The station number in (B) corresponds to the station index in (A). Triangles denote seismic stations, and two malfunctioning stations are marked in gray. Black stars represent the optimal locations determined by our system, and white stars mark the cataloged locations.

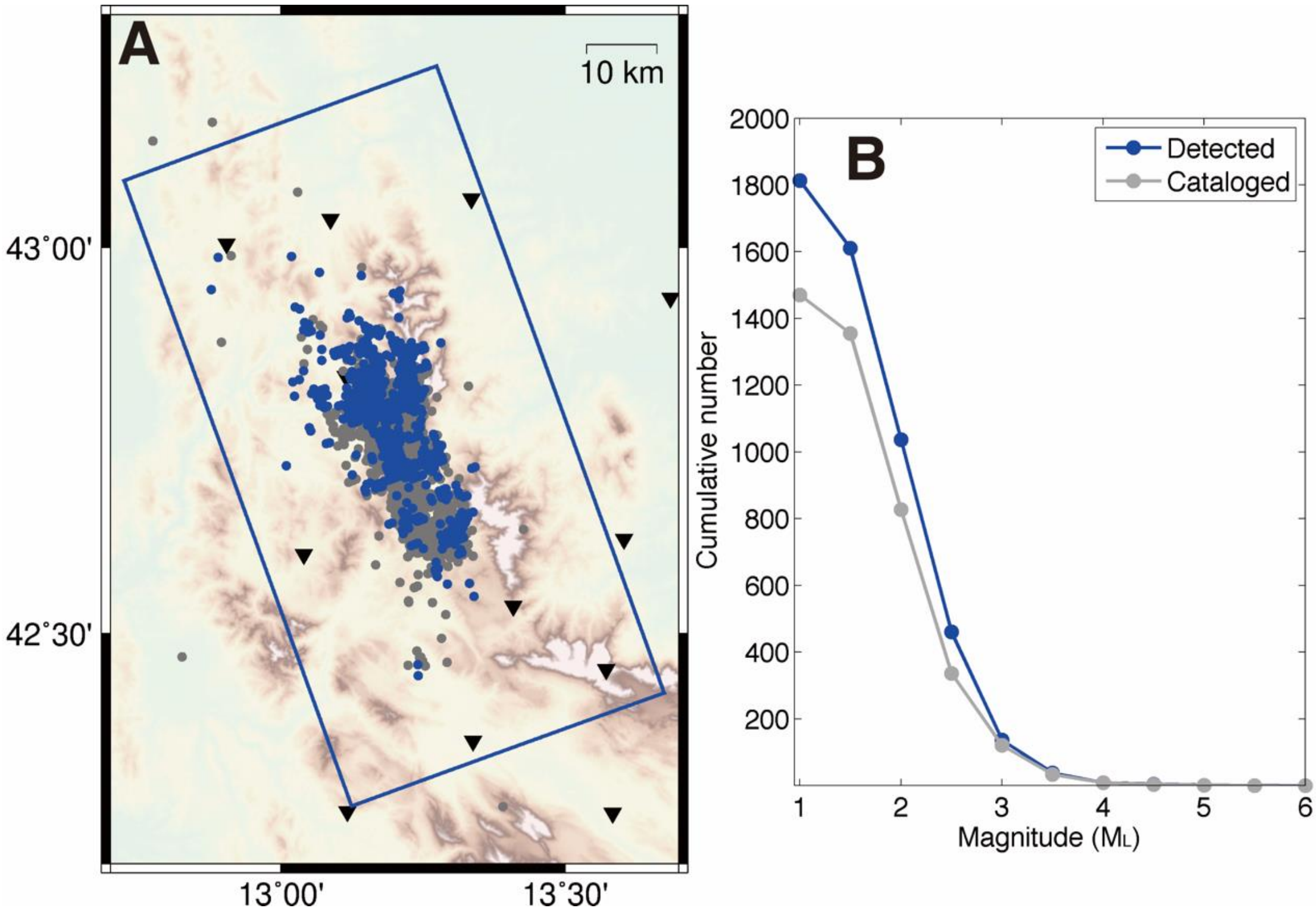

**Fig. 3. Source parameter comparison for earthquakes on August 24, 2016 between our detection and the catalog.** (A) Epicentral location and (B) magnitude. Blue and gray dots represent earthquakes detected by the system and cataloged events, respectively. Only M≥1.0 events were used for magnitude comparison.

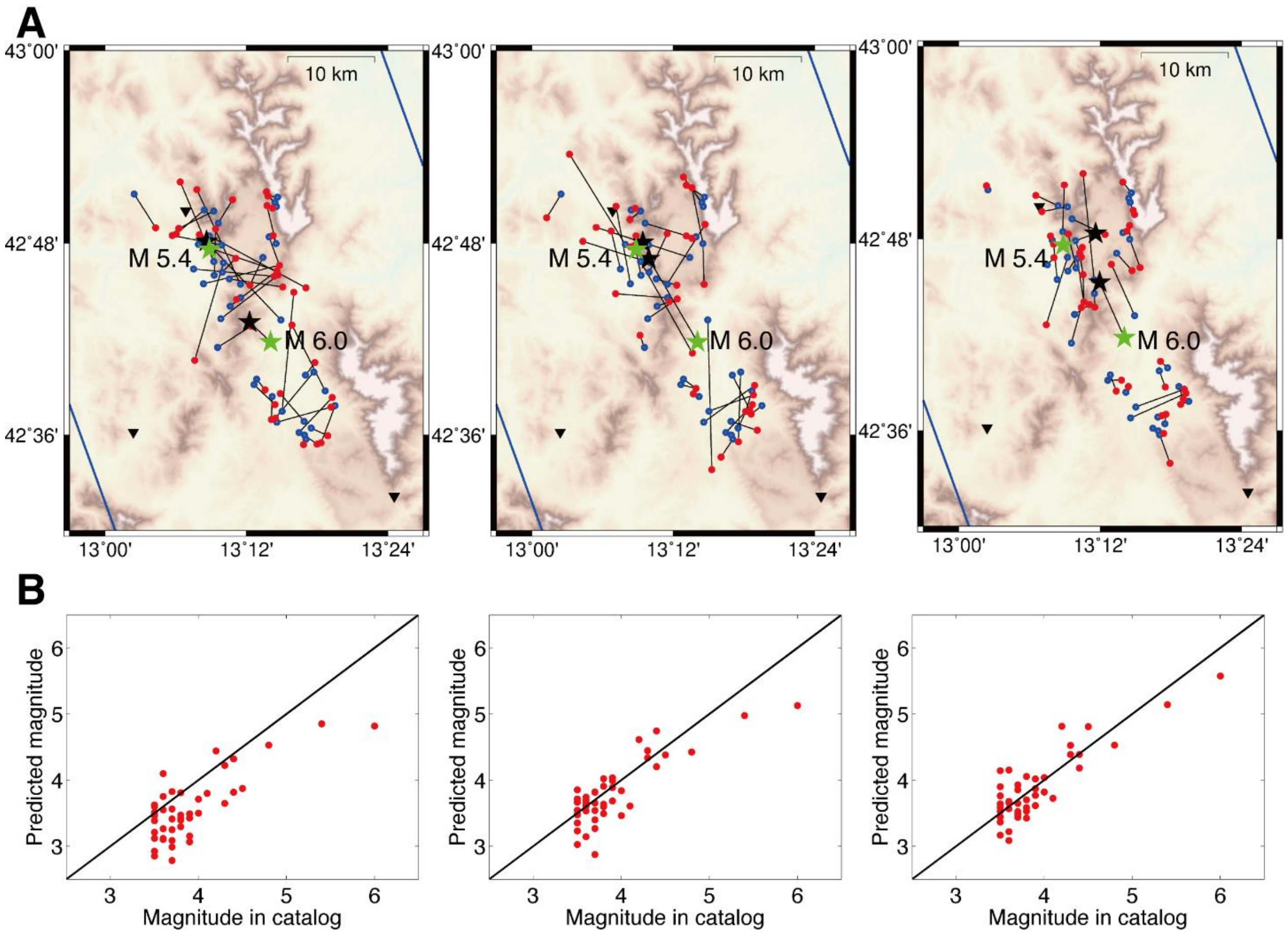

**Fig. 4**. **Source parameter comparison for the 43 M≥3.5 aftershocks between our results and the catalog.** (A) Epicentral location and (B) magnitude. Three snapshots are shown at 4 s, 9 s, and 15 s after the first P arrival (from left to right). Locations of common earthquakes are linked together between our results (red dots) and the catalog (blue dots). Black stars and green stars denote epicentral locations of the two M>5.0 earthquakes in our results and the catalog, respectively. Note that the magnitudes of the two M>5.0 earthquakes are underestimated because of waveform clipping.

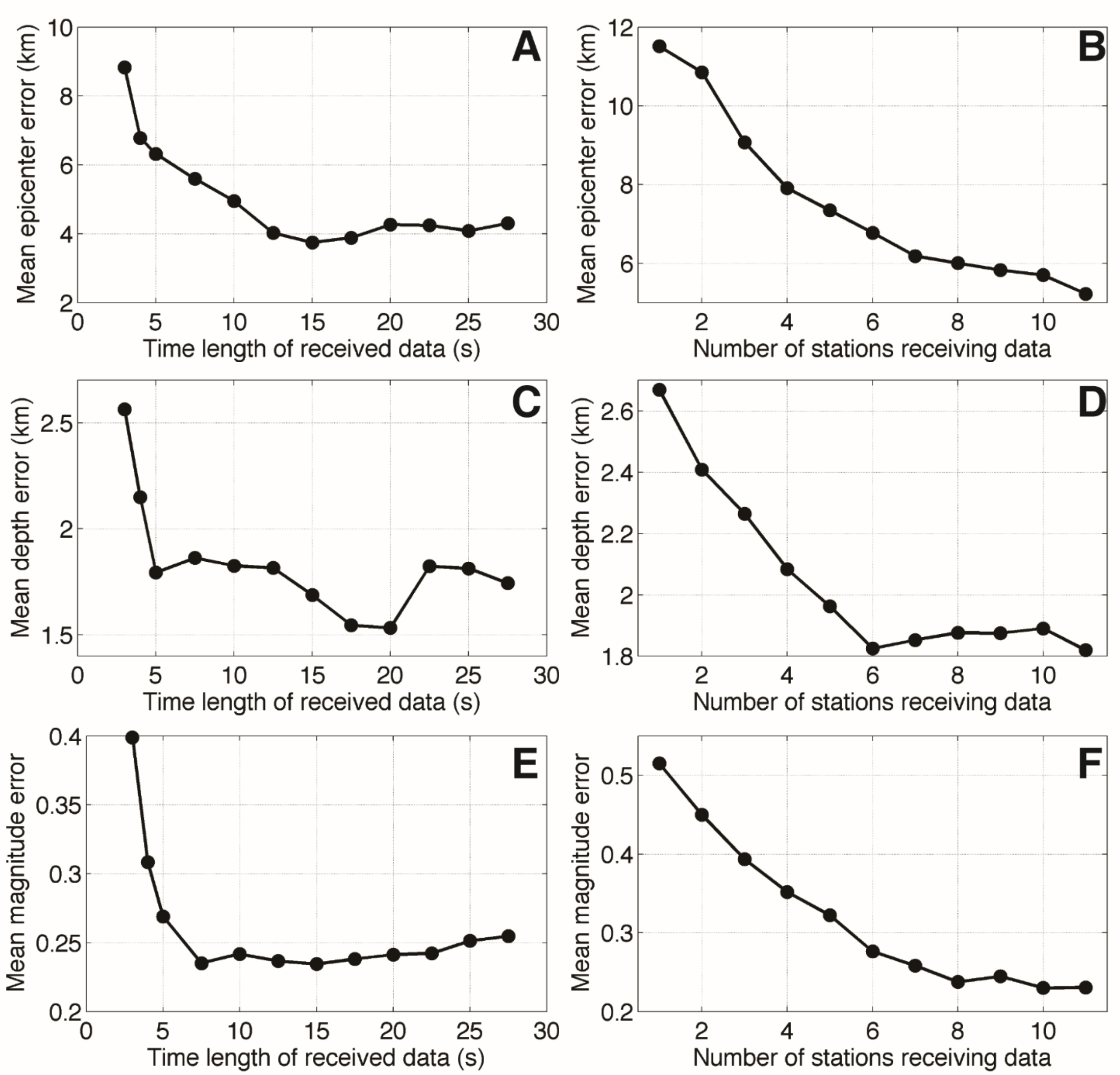

**Fig. 5**. **Error analysis of source parameters for 500 testing earthquakes.** Mean errors of (A, B) epicentral location, (C, D) depth, and (E, F) magnitude change with the time length of the received data and the number of stations receiving data.

## DISCUSSION

### Fully convolutional network

By utilizing the full convolutional network (FCN), we adopt a Gaussian distribution to label the earthquake location and magnitude parameter, enabling us to exact source parameter values rather than the categories to which the parameters belong. Although the large number of trainable parameters potentially cause over-parameterization, deep neural networks are helpful for improving network generalization and training convergence (*31*-*32*). The overfitting problem in deep learning is eliminated by dropout layers (*27*). The FCN method was fully tested and analyzed by Zhang et al. (2020) for earthquake location. In this study, we focus on its performance in real-time earthquake monitoring for early warning.

### Advantages compared with conventional EEW systems

Our deep learning EEW system shows significant advantages in terms of both efficiency and reliability compared with conventional EEW systems. First, we directly extract earthquake source parameters from continuous waveform streams without phase picking, which substantially decreases the response time for early warning. Second, with the “the triggered and not-yet-triggered stations" concept (*15-16*), both earthquake signals and noises are utilized for constraining earthquake location in the neural network. Third, deep learning neural networks are not sensitive to abnormal waveforms (*19*), which potentially reduces the possibility of false alarms. Finally, we are able to detect events and determine their source parameters as early as a few seconds after the first P phase at one or two stations and continue updating the source information in real time as more data is received. This allows for a better understanding of earthquakes over time.

### Unbalanced magnitude distribution in the training set

The percentage of small earthquakes is much greater than that of large earthquakes in the training set because specific regions have many more small events (Fig. S1). This problem potentially affects both location and magnitude estimations in deep-learning-based methods. Waveform features in large and small earthquakes may be different because of their distinct rupture scales (i.e., point source vs. finite fault). For location, our neural network is able to handle large events with an acceptable error range, but the mean error is relatively larger than the average. Other studies point out more serious issues with the magnitude estimation (*24, 33*). To eliminate this problem, we introduce a normalized magnitude $M_r$ to label earthquake samples by excluding the contribution of absolute amplitudes and equalizing the labeled magnitude range between large and small earthquakes. With the exception of the M 6.0 mainshock and M 5.4 aftershock that were affected by waveform clipping, our predicted magnitudes for large events are as good as those for small events; waveform clipping can be addressed by utilizing strong motion recordings if available. While the location of earthquakes by the neural network can still be improved by increasing the percentage of large events in the training set by, for instance, generating synthetic waveforms of large events. However, this is challenged by accurate 3D velocity models and high-frequency waveform simulation.

### Potential improvements

We adopted cataloged earthquakes for our neural network training and testing. However, they are not ground truth and inevitably contain uncertainties regarding the location and magnitude, which may originate from phase picking and velocity models and from the algorithms themselves. Thus, the network model may be further improved by adopting high-precision earthquake relocations for network training (*34*). Additionally, the monitoring ability and accuracy are expected to be improved by upgrading seismic station coverage.

In our system, the earthquake origin time is obtained based on the travel time difference between the earliest available P phase arrival and the source location. Our neural network automatically outputs the earthquake location when an earthquake is detected; then, the first P phase is picked by a deep learning picker—PhaseNet—within the current time window (*19*). A more straightforward way is to design another branch of the neural network to directly output the origin time, which will be tested in our future work.

Results show that our neural network generally works well for real-time EEW except for two extreme cases: 1) the system may miss detection or result in false detection when waveforms overlay with coda waves of large earthquakes; 2) the system may not distinguish multiple events when they occur closely one after another within a short time period (e.g., 30 sec), potentially leading to event missing or source parameter mixing up between the events. Both cases are challenging in available EEW algorithms as well. But we may join forces with various techniques to improve the warning reliability in practical EEW.

EEWs are generally expected to be improved by 1) developing rapid algorithms to reduce response time and 2) improving seismic station coverage to receive earthquake signals as early as possible. We focus on the first strategy to develop a deep learning EEW system to provide earthquake source parameter information in real time. Recent studies suggest that smartphones can significantly improve seismic data coverage for earthquake early warning in high-population regions (*35-36*). A future breakthrough may lie in the combination of algorithm development and the improvement of station/sensor coverage.

### Conclusion

We propose a deep-learning EEW system for providing earthquake source parameters in real time. Without phase picking, our deep learning neural network directly extracts earthquake location and magnitude information from seismic waveform streams and evolutionarily updates the solutions. Our system could provide reliable earthquake source parameters within a few seconds of effective signals received at one or two stations. We successfully applied the system to the 2016 Central Apennines, Italy mainshock and its subsequent aftershocks. Our results demonstrate the feasibility of using deep learning techniques for real-time earthquake early warning.

## METHODS

We designed a multi-branch neural network for earthquake location and magnitude estimation. We adapt the earthquake locations and magnitudes to the application of the FCN with 3D and 1D Gaussian distribution probability labels, respectively, instead of assigning exact values (Fig. 1). The labels are centered at the cataloged earthquake locations and normalized magnitudes (see later definition). For earthquake location, inputs of our neural network are three-component waveforms from the 12 permanent broadband seismic stations, with a size of 2048 (time samples) × 12 (number of stations) × 3 (number of components). We discretize the study region and set the grid interval as 0.5 km in the horizontal and vertical directions; thus, the output for the location branch of our neural network is a 3D matrix (192 × 96 × 64) representing the monitoring region (i.e., length: 96 km, width: 48 km, depth: 32 km). For magnitude estimation, we define a normalized magnitude $M_r = M - \log(A_{max})$ by deducing the waveform amplitude ($A_{max}$) contribution, which enables us to utilize the neural network to solve for magnitude with normalized waveforms. Thus, $M_r$ is independent of magnitude and represents the constant correction term and the falloff curve with the distance (*37*). In addition to three-component waveforms, we need to keep the maximum amplitude among stations for both the training and testing samples to convert the normalized magnitude $M_r$ to the true magnitude M. We label normalized magnitudes $M_r$ as 1D Gaussian distribution probabilities centered at the normalized magnitudes of training earthquakes. In the monitoring stage, the neural network outputs a 3D probability distribution for the location of an earthquake and a 1D probability distribution for its normalized magnitude $M_r$. If the maximum location probability is larger than our preset empirical threshold (0.9), an earthquake is detected, and the optimal location and normalized magnitude $M_r$ are determined according to the positions of their maximum probabilities. By memorizing the maximum amplitude of the input waveforms, we convert the normalized magnitude $M_r$ to the true magnitude *M*. We apply a deep learning phase picker—PhaseNet—to pick the earliest P phase within the current time window (*19*). The origin time is estimated by calculating the travel time difference between the station and the earthquake location. Therefore, the system simultaneously determines the location and the magnitude, as well as the origin time.

**Network architecture**

The input layers are connected to two branches of the neural network representing earthquake location and magnitude estimation. For the location branch, similar to the network architecture proposed by Zhang et al. (2020), the channel number increases to 1024 and then decreases to 64 through 28 convolutional layers. The max-pooling and upsampling layers adjust the feature size and transform the input size (12 × 2048) to be the location size (96 × 192). We first utilize the four max-pooling layers to downsample the input features. Then, we increase the feature size by using three upsampling layers. Therefore, the final output size is 96 × 192 × 64, corresponding to a study area of 48 × 96 km with a depth range of 32 km. For the magnitude branch, the structure is similar to that of the location branch. The channel number increases to 1024, but the final channel is one. The four maxpooling and three upsampling layers transform the input size (12 × 2048) to the magnitude of the image size (1 × 512). Table 1 shows the detailed settings of the neural network.

**Source parameter labeling**

We prepare the training samples by labeling the truncated waveform data with Gaussian distribution probabilities for locations and magnitudes. The location and magnitude probabilities are projected as follows:

$$\begin{cases} f(x,y,z)=\exp\left\{-\left[(x-x_0)^2+(y-y_0)^2+(z-z_0)^2\right]/r\right\} \\ x\in R_x, y\in R_y, z\in R_z \\ g(M_r)=\exp\left[-(M_r-M_{r0})^2/r'\right] \\ M_r\in R_{M_r} \end{cases} ,(1)$$

where $(x_0,y_0,z_0)$ and $M_{r0}$ denote the true earthquake location and normalized magnitude, $r$ and $r'$ are the radius of the Gaussian functions for location and magnitude, respectively; $R_x$, $R_y$, $R_z$, and $R_{M_r}$ are ranges of locations and normalized magnitudes. We set the radius parameters $r$ and $r'$ to 25 (i.e., 5 km for location) and 0.16 (i.e., 0.4 for magnitude), and calculated the Gaussian distribution value (*f* and *g*) at each grid node of the location and magnitude.

**Objective function**

The prediction function is defined by optimizing the cross-entropy loss function as follows:

$$\Psi=\frac{1}{N}\sum_{k=1}^{N}\sum_{d\in D}p_d^k\log\left(q_d^k\right)+\left(1-p_d^k\right)\log\left(1-q_d^k\right) , \quad (2)$$

where *p* and *q* are the predicted and true location probability labels, *N* is the number of training samples, and *D* is the assemblage of the grid nodes. The loss function measures the difference between the predicted and true probability distributions. We utilize the same loss function to simultaneously constrain the location and magnitude, feed the neural network with training samples, and minimize the loss function to obtain the final neural network model. The ADAM algorithm is utilized to optimize the loss function (*38*) with defaulted parameter values recommended by the authors—except the set learning rate of $10^{-4}$. To prevent overfitting, we introduce two dropout layers by setting the dropout rate to 0.5 for each branch of the network.

**Training the network**

We trained our neural network on Tensorflow (39) before applying it to continuous waveform streams. For the earthquake location, waveforms were filtered from 2–8 Hz. For the magnitude estimation, we filter waveforms in the range of 0.5–9 Hz and keep the maximum amplitude $A_{max}$ among stations. Additionally, to distinguish seismic noises from effective seismic signals, we add 60 noise samples to the training set and label their location probabilities as zeroes. We apply the widely used data augmentation strategy to generalize our neural network by truncating the waveforms of training samples in different time windows. We randomly cut waveforms of training samples five times from 0.5–27 s relative to their first theoretical P phase with a total length of 30 s to cover various cases with few stations containing earthquake signals in truncated time windows. Thus, the total number of training samples is 12,590, consisting of 12,530 earthquake samples (i.e., 2506 × 5) and 60 noise samples. We utilized 200 epochs with a batch size of four to train the network and select the model with the smallest validation loss as our final neural network model. We selected 15% of the training set as validation samples and adopted the model with the lowest validation loss. Our well-trained neural network is able to immediately respond to earthquake location and magnitude if effective seismic signals are recorded at one station. The network evolutionarily updates earthquake source parameters by receiving more waveforms recorded at one or multiple stations.

**Acknowledgments: Founding:** This work was supported by the National Natural Science Foundation of China Grant (No. 41704040), National Science Foundation Grant (No. EAR-1759810), and the Natural Science and Engineering Research Council of Canada Discovery Grant (No. RGPIN-2019-04297). **Author contributions:** X. Z. developed the method, processed the data, and analyzed the results. M. Z. designed the project and analyzed the results. X. T. helped to process the data. X. Z., M. Z., and X. T. wrote the manuscript. All authors contributed to the project. **Competing interests**: The authors have no competing interests. **Data and materials availability**: The maps in our paper were made using Generic Mapping Tools (*40*) and MATLAB(https://www.mathworks.com/products/matlab.html). Seismic data were downloaded from Italy's National Institute of Geophysics and Volcanology (INGV) through the International Federation of Digital Seismograph Networks (FDSN) web services. The earthquake catalog used in this study can be downloaded from http://cnt.rm.ingv.it/ (last accessed May 2020).

## Supplementary Materials

### Contents of this file

Figures S1 to S6

### Additional Supplementary Materials (uploaded separately)

Movie S1. Real-time monitoring of the M 6.0 mainshock and its first hour aftershocks.
Available at:
https://drive.google.com/file/d/1BLNyv17B1G-Cg-jSOSHyNExDRsQJ9lLm/view

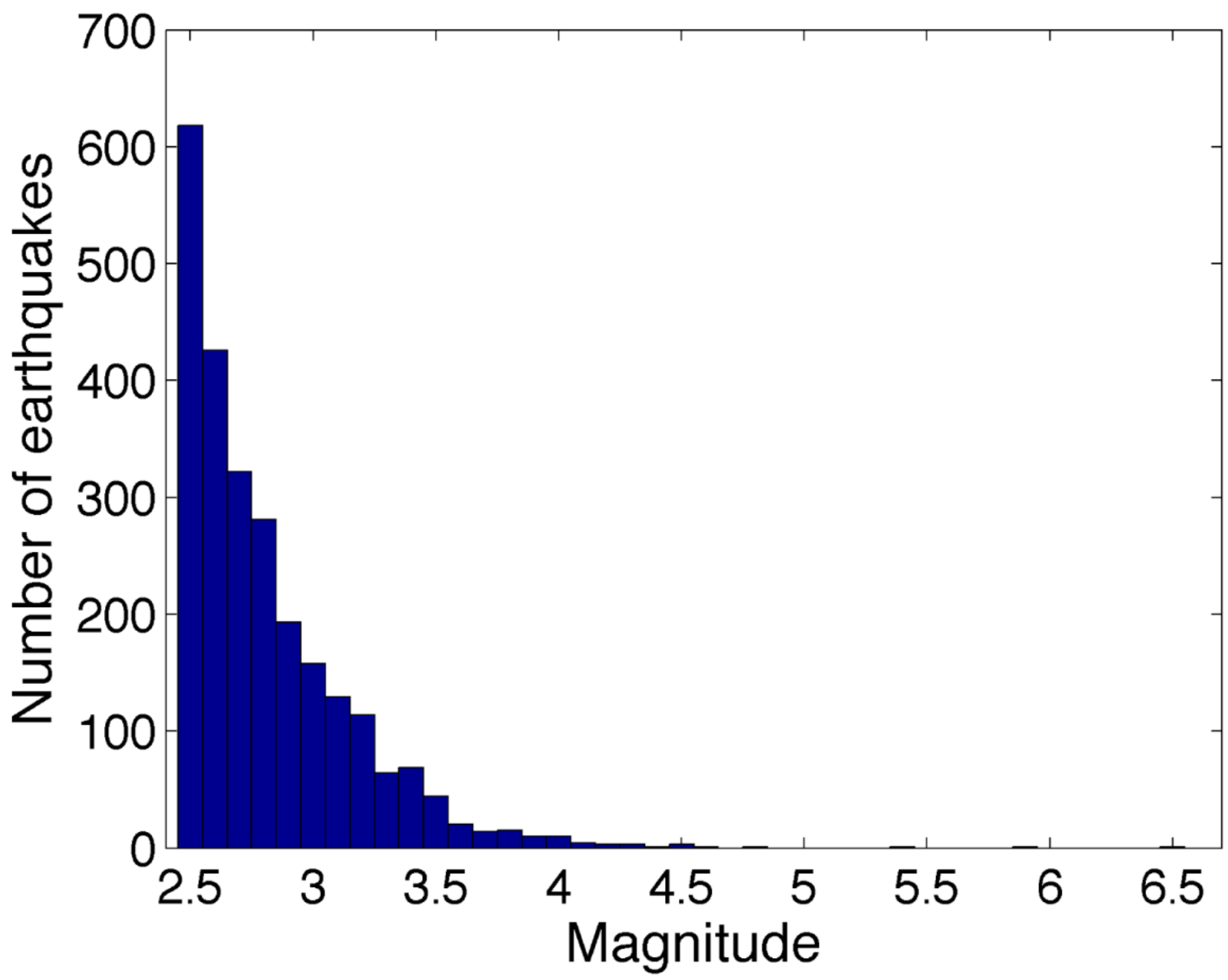

**Fig. S1. Magnitude distribution for training earthquakes.**

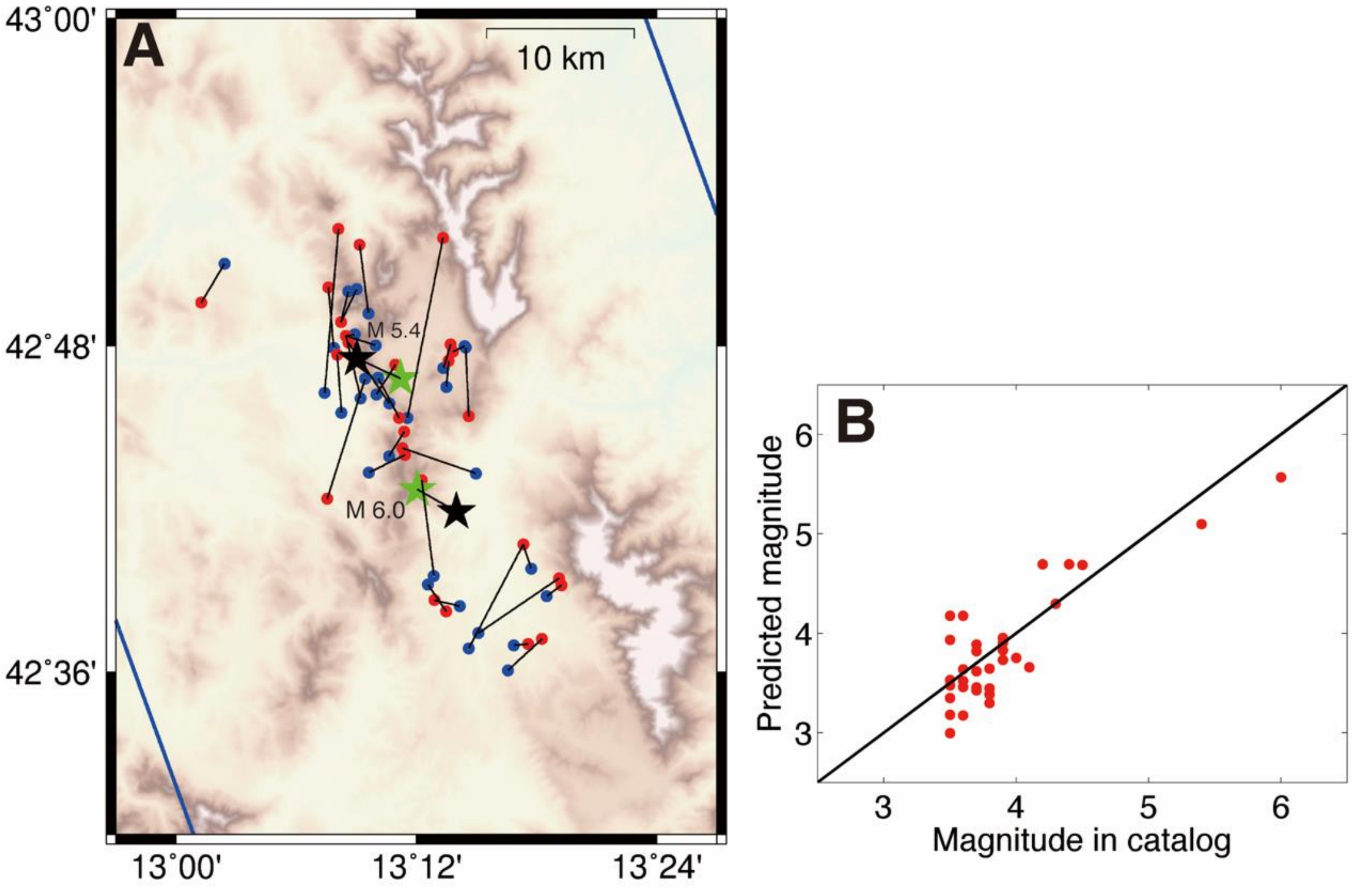

**Fig. S2. Source parameter comparison for the 33 M≥3.5 common events on August 24, 2016 between our results and the catalog.** (A) Epicentral location and (B) magnitude. Locations of common earthquakes are linked together between our results (red dots) and the catalog (blue dots). Green stars and black stars denote epicentral locations of the two M>5.0 earthquakes in our results and the catalog, respectively. Note that the magnitudes of the two M>5.0 earthquakes are underestimated because of waveform clipping.

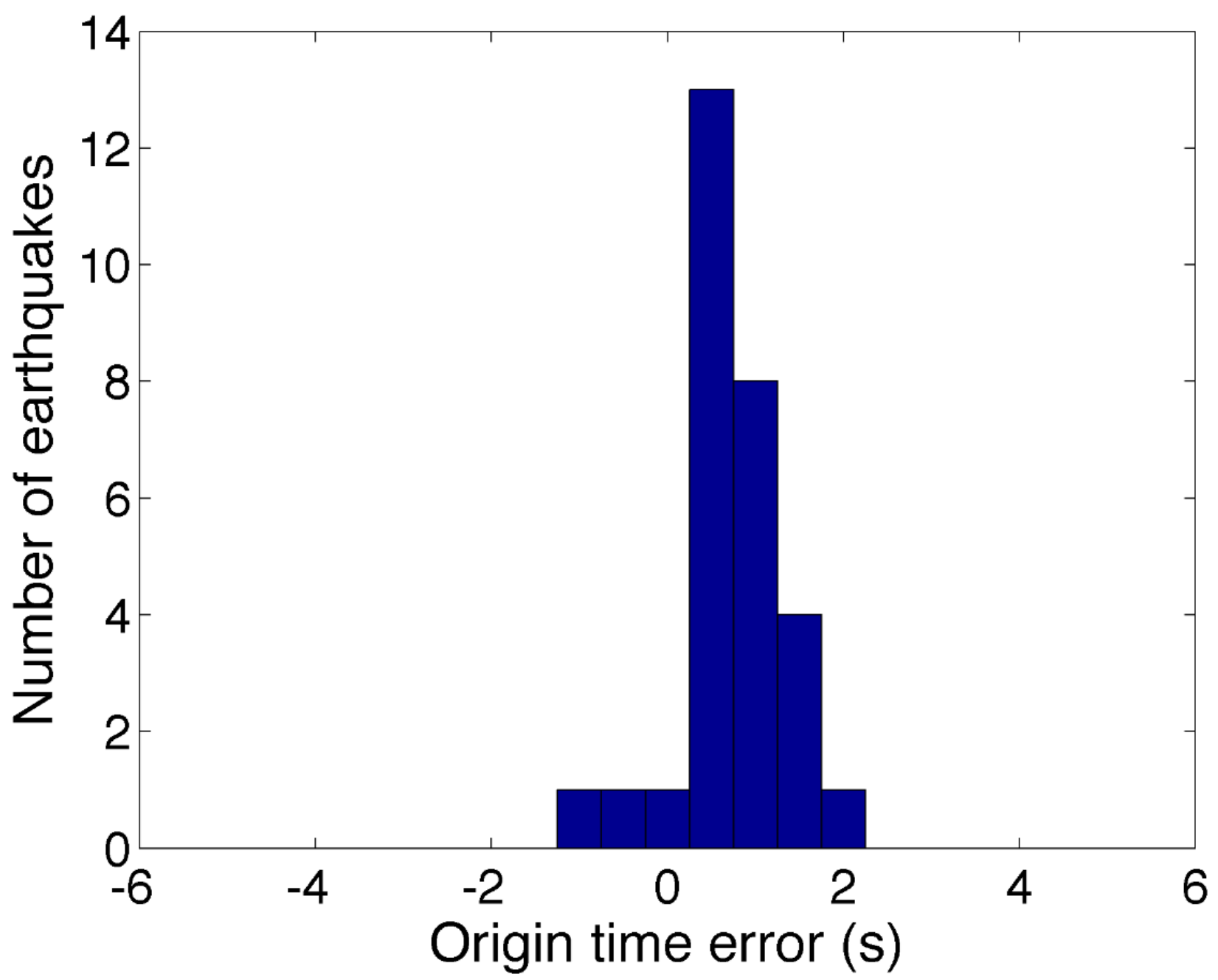

**Fig.S3. Origin time error comparison for the 31 common events with origin time errors less than 2 s between our result and the catalog.**

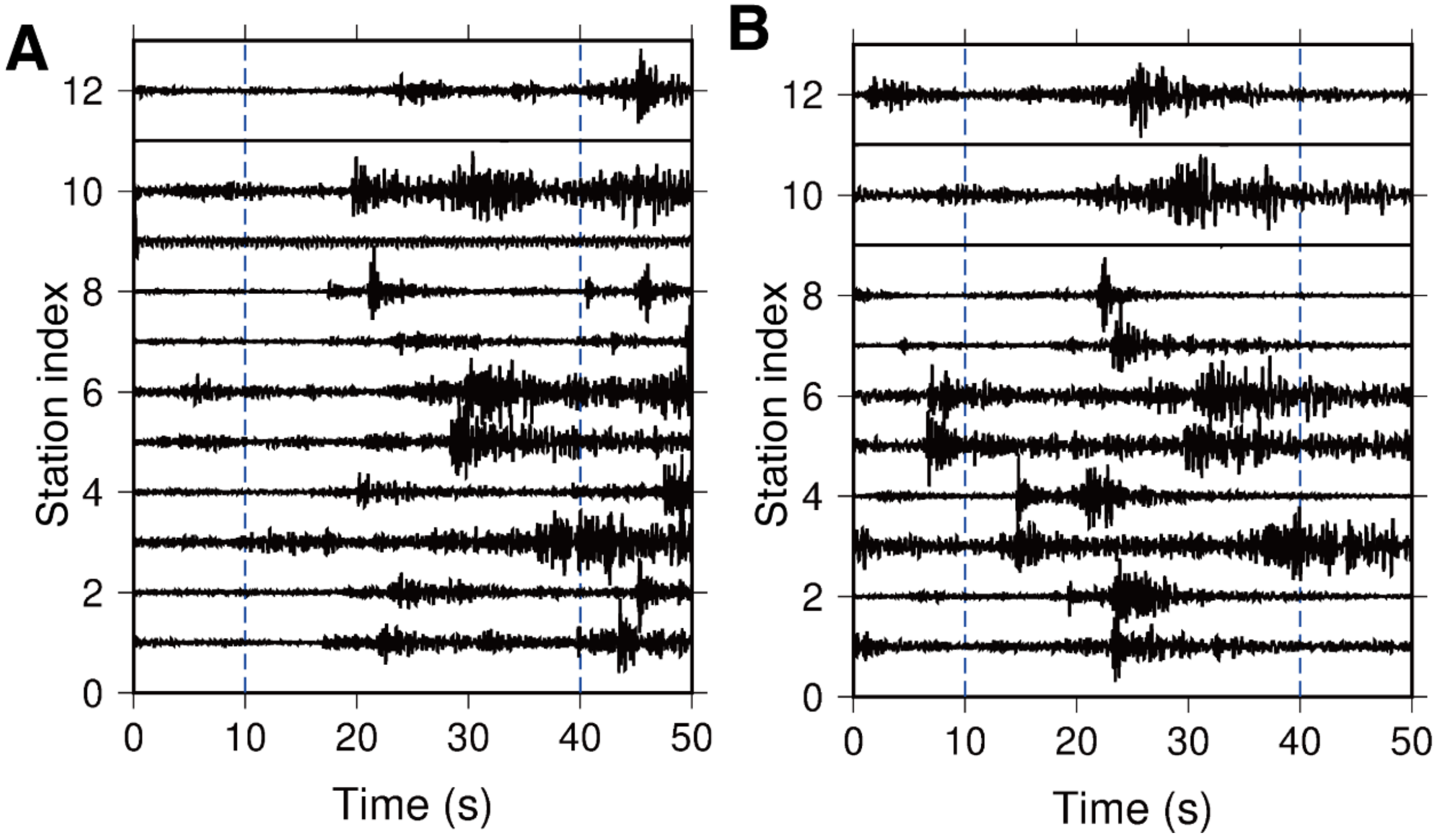

**Fig. S4: Two earthquakes with large origin time error due to waveform overlapping with other events.** Corresponding time windows (10 – 40 s): (A) 20160824T01:55:20 and (B) 20160824T02:12:11.

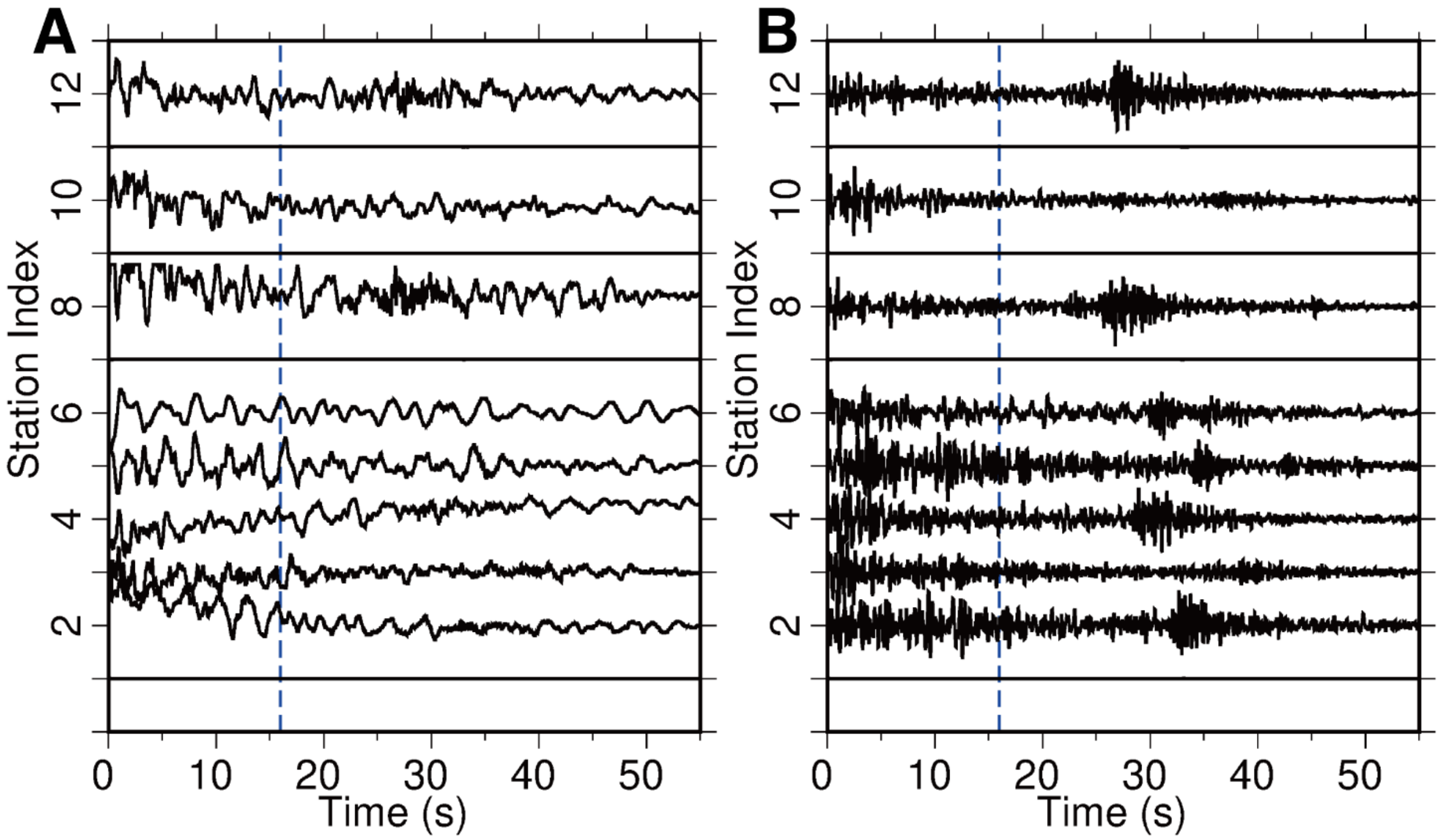

**Fig. S5. Waveforms of the missed earthquake occurred following the M 6.0 mainshock.** (A) Raw and (B) 2–8 Hz filtered waveforms. Blue dash lines denotes its cataloged origin time.

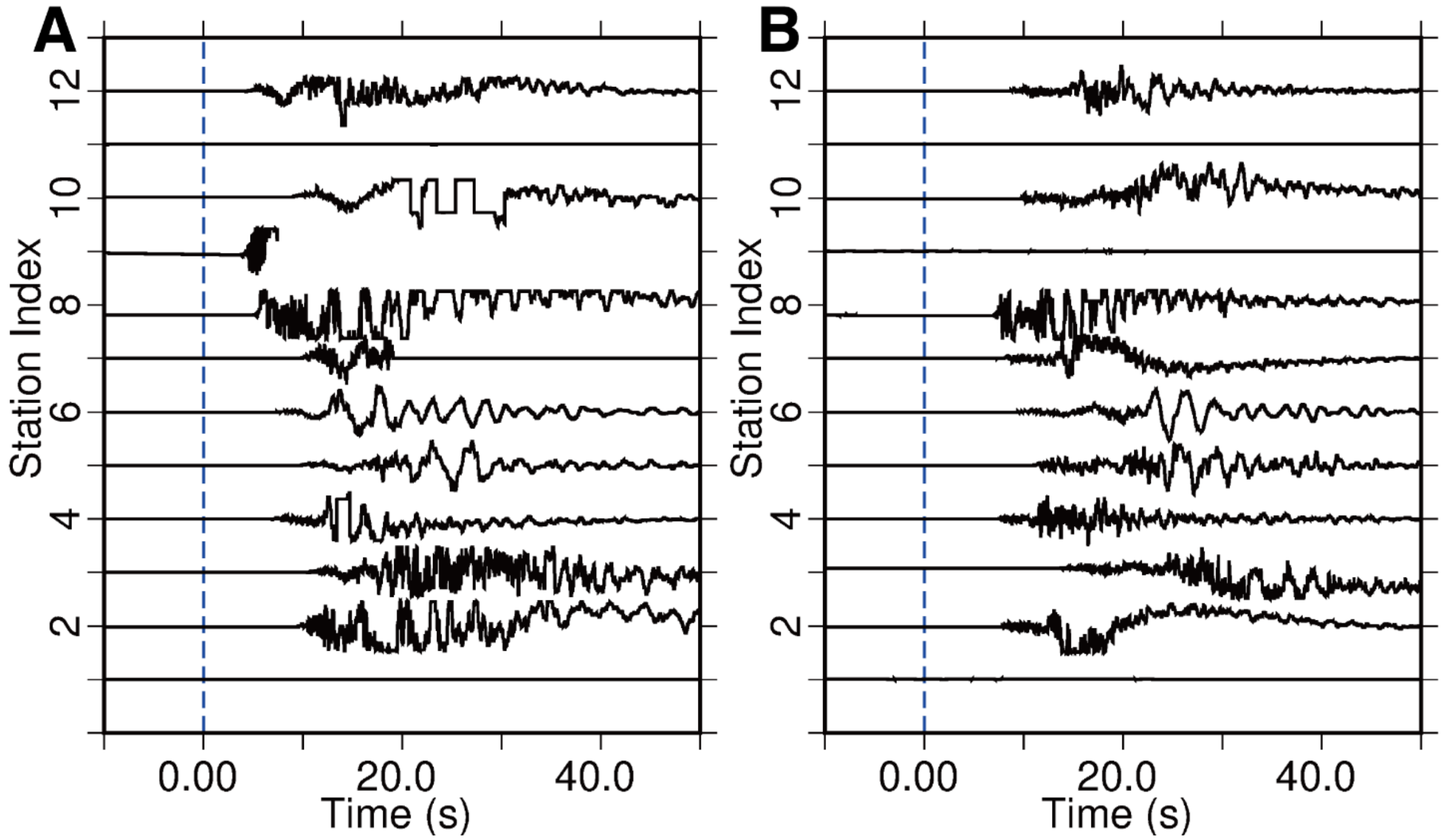

**Fig. S6. Self-normalized vertical raw waveforms for the two M>5 earthquakes.** (A) M 6.0 mainshock and (B) the M 5.4 aftershock. Clearly, both waveforms were clipped. Blue dash lines denote their cataloged origin time.